\documentclass[iop]{emulateapj}
\shortauthors{Ben-Jaffel \& Holberg}
\shorttitle{Voyager UVS calibration \& the heliosphere}
\newcommand{\lya}{Lyman-$\alpha$} 
\newcommand{\lyb}{Lyman-$\beta$} 
\usepackage{natbib}
\begin{document}
\title{Voyager Ultraviolet Spectrometers calibration and the heliosphere neutrals composition: reassessment}
\author{Lotfi Ben-Jaffel \altaffilmark{1}}
\affil{Sorbonne UniversitŽ\'es, UPMC Univ Paris 6 \& CNRS, UMR 7095, Institut d'Astrophysique de Paris, 98 bis bd Arago, 75014 Paris, France}
\email{bjaffel@iap.fr}
\altaffiltext{1}{Visiting Research Scientist, University of Arizona, Lunar \& Planetary Laboratory, Tucson, USA}
\and
\author{J. B. Holberg}
\affil{University of Arizona, Lunar \& Planetary Laboratory, Tucson, USA}
\email{holberg@vega.lpl.arizona.edu}
\begin{abstract}
{The Voyagers (V) 1 and 2 Ultraviolet Spectrometers (UVS) data harvest covers outer planets encounters, heliosphere sky-background measurements, and stellar spectrophotometry. Because their operation period overlaps with many ultraviolet missions, the V1 and V2 UVS calibration with other spectrometers are invaluable. Here we revisit the UVS calibration to assess the intriguing 243\% (V1) and 156\% (V2) sensitivity enhancements recently proposed. Using the Saturn \lya\, airglow, observed {in-situ} by both Voyagers, and remotely by IUE, we match the Voyager values to IUE, taking into account the shape of the Saturn \lya\, line observed with the Goddard High Resolution Spectrograph onboard the Hubble Space Telescope. For all known ranges of the interplanetary hydrogen density, we show that the V1 and V2 UVS sensitivities cannot be enhanced by the amounts thus far proposed. The same diagnostic holds for distinct channels covering the diffuse HeI 58.4\,nm emission. Our prescription is to keep the original calibration of the Voyager UVS with a maximum uncertainty of 30\%, making both instruments some of the most stable EUV/FUV spectrographs of the history of space exploration. In that frame, we reassess the \lya\, emission excess detected by Voyager UVS deep in the heliosphere, to show its consistency with the heliospheric but not the galactic origin. Our finding confirms results obtained nearly two decades ago--namely, the UVS discovery of the heliosphere distortion and the corresponding local interstellar magnetic field's obliquity ($\sim40^{\circ}$ from upwind) in the solar system neighborhood-- without requiring any revision of the Voyager UVS calibration.} 
\end{abstract}
\keywords{interplanetary medium---radiative transfer---ultraviolet: general---instrumentation: spectrographs---Sun: heliosphere---ISM: atoms}

\section{Introduction} 

The far and extreme ultraviolet (FUV \& EUV) wavelength windows contain most of the neutral and ions species'  atomic and molecular lines and bands. The access to neutral hydrogen and helium distributions in planetary atmospheres and the heliosphere are key ingredients, respectively, to understand the formation and evolution of the atmospheric envelope of planets and to derive the helium content in the local interstellar medium \citep{ben15}. The paradigm thus far assumed is that planetary airglow and sky background diffuse emissions could be converted to species abundances as far as the radiative processes and the photons sources are constrained. However, this approach requires taking into account inherent problems regarding radiation transfer (RT) effects in moving and very extended, optically thick media, along with difficulties in accurately monitoring over time key parameters like the solar wind, the solar \lya\ flux, or the calibration of the instruments that measure the FUV and EUV airglow. While most problems could be addressed using sophisticated modeling and comparison to observations, the instrument calibration remains a difficult problem, particularly the absolute calibration and the inevitable time degradation of the apparatus over the lifetime of a space mission.

The Ultraviolet Spectrometers (UVS) aboard the Voyagers 1 and 2 spacecraft have been operating in space since late 1977\footnote{The V2 UVS was turned off in late 1998 to conserve spacecraft power}. The UVS observe both diffuse and point-like sources, which include planetary airglow, sky-background emissions, and key stellar targets. The two instruments, which cover the 50-170\,nm spectral range, were independently calibrated in the laboratory and in-flight \citep{bro77,bro81}. The corresponding calibration pipeline is well documented in \citet{bro81,hol92}\footnote{http://vega.lpl.arizona.edu/Voyager{$\_$}Ultraviolet{$\_$}Spectrometer.pdf}. The Voyager UVS instruments have a unique ability to observe the spectra of hot stars from the Lyman limit at 91.2\,nm to 170\,nm, and in the case of a few hot white dwarfs, also below the Lyman limit.  In comparing the Voyager absolute stellar fluxes with published fluxes obtained from sounding rockets and from the International Ultraviolet Explorer (IUE), it was noted that although the two Voyagers agreed with one another, the fluxes did not agree with published values in the sense that the Voyager fluxes were too high. This issue was studied carefully by \citet{hol82} and the following conclusions were reached: (1) In the wavelength range longward of \lya\, all observations agreed. (2) Between 91.2\,nm and 115\,nm all observations disagreed, sometimes by as much as a factor of three.  For a number of reasons elaborated in \citet{hol91}, the decision was made to reduce Voyager 2 fluxes by a factor of 1.6, based on a comparison with model atmosphere fluxes from the hot white dwarf HZ\,43.  The Voyager calibration was left unchanged at 58.4\,nm and was linearly interpolated between 58.4\,nm and 91.2\,nm.  Voyager 1 fluxes were adjusted to agree with Voyager 2.  These changes, made in 1982, still constitute current Voyager stellar UV and EUV calibration. The issue of Voyager calibration was revisited in \citet{hol91}, which provided additional compelling evidence from very hot subdwarf stars that the calibration was essentially correct.  Since that time, numerous results have been published comparing Voyager stellar observations with those from the Far Ultraviolet Spectroscopic Explorer (FUSE) and the Hopkins Ultraviolet Telescope (HUT) where agreement is at the 10 \% level \citep{kru97}.

During the Voyager 1 Jupiter encounter in 1979, the UVS endured excessive radiation induced counting from $> 3$ Mev electrons in the inner Jovian magnetosphere. The result was that post-Jupiter V1 spectra showed a marked difference with respect to pre-Jupiter spectra. These changes manifest themselves in two ways: a reduced response across the detector, and a change in the channel-to-channel relative response. The explanation for these changes was a "gain-sag" in the microchannel plate caused by the excessive counts that effectively reduced detector gain.  Since each individual channel had an independent readout and set of upper- and lower-level voltage discriminators, the channel-to-channel responses changed in response to the change in gain, giving the spectra an uncharacteristic noisy appearance. To remedy this situation, a careful study of pre- and post-Jupiter spectra was made to define a new fixed-pattern-noise (fpn) vector to correct the post-Jupiter spectra to their pre-Jupiter appearance and level \citep{hol82}. This correction was applied to all subsequent Voyager 1 spectra.  Note that following the Jupiter encounter, some reports applied the correction on either the full UVS spectral range including the He 58.4\,nm line \citep{san82} or on wavelengths larger than 92\,nm \citep{she83}, which added to the confusion about instrument calibration. Note also that Voyager 2 avoided this problem by reducing detector voltages during the critical passage through the inner Jovian magnetosphere. These changes to Voyager 1 were the only documented changes to the UVS instruments over the entire mission.

The calibration of the \lya\, region (120 -129)\,nm requires some additional discussion. The adjustments to the UVS calibration discussed above were based on stellar spectra.  However, in all cases the stellar profiles were relatively broad and nearly saturated features, which made comparisons of Voyager flux within this spectral range difficult. For instance, to estimate the reduced response across the V1 detector after the Jupiter flyby, two independent analyses used reference to stellar spectra (respectively Alpha Virgo and Alpha Leo) recorded before and after the encounter to derive a $(32 \pm 7)$\% degradation in the V1 UVS effective sensitivity in the spectral windows (100-119)\,nm and (130-145)\,nm \citep{hol91,hal92}. The degradation was also applied to \lya\, channels, but for a very short period of time around the V1-Saturn encounter. Soon afterward, an independent study used the light scattered in the instrument to test whether changes in sensitivity could be detected. As shown in \citet[][see their page 171, Figure A.1]{hal92}, a signature of \lya\, instrumental scattering is the formation of a faint, spectrally-extended emission feature around the base of the line core. By comparing channels corresponding to the line core (70-78) and adjacent channels covering the extended scattering feature before and after the encounter of Jupiter, \citet{hal92} could show that the \lya\, channels suffered $(0.67 \pm 0.07)$ less degradation than the adjacent channels used for the stellar calibration  \citep{hal92}. According to \citet{hal92}, taken together, the two later-proposed corrections cancel each other, which led to the significant conclusion that the \lya\, channel sensitivity remained unchanged for V1 before and after the Jupiter encounter. For consistency, \citet{hal92} also proposed:  
\begin{itemize}
\item first, to use the V2 UVS pre-flight laboratory calibration curve to generate a synthetic spectrum that is compared to a high-quality sky background line, which should provide an absolute calibration for V2 UVS;
\item to use the relative ratio between V1 and V2 UVS sensitivities obtained after the Jupiter encounter;
\item and finally, to derive the V1 UVS sensitivity for the post-Jupiter encounter period.
\end{itemize}

The analysis reported in Hall's thesis led to \lya\, absolute calibrations of $218\pm33$\, R/counts/s (in channels 70-78) for V1 and 172 R/counts/s (in channels 72-80) for V2, with a ratio $(0.79\pm 0.12)$ between the two instruments that was derived for the post-Jupiter period. It is significant that the \citet{hal92} calibration at \lya\, is independent of any RT modeling of the sky background and of any solar \lya\, flux assumed (see next section). The consistency between the pre-flight and post-flight calibration, in addition to the inter-calibration agreement between V1 and V2, represent the strength of the calibration process proposed by \citet{hal92}.

Recently, \citet{que13} used the radial distribution of the sky-background \lya\, emission recorded by V1 and V2 along their trajectories in the heliosphere to derive a new, independent estimation of the UVS instruments flux calibration. Those authors carefully emphasize that the proposed calibration was the first step in a complex approach aiming to inter-calibrate most of existing UV space missions, with the sky background as the common target to bring the different instruments into consistent, absolute calibrations over space and time. They made use of sophisticated three-dimensional kinetic models of the heliospheric H\,I distribution, including charge exchange with the local plasma, as well as 3D RT models to describe the transport of the solar \lya\, photons inside a very extended and moving medium, which is bound by the Sun inside and the local interstellar cloud (LIC) outside \citep{que13}. From the comparison between their model and UVS observations, Qu\'emerais et al. conclude that there is a need to correct the V1 UVS sensitivity by a large factor of 243\% and the V2 UVS by a factor of 156\% with respect to the calibration thus far proposed by \citet{hal92}. The authors stress that the new calibration is based on a fit to the sky background observed by V1 and V2 UVS in the region { 10-50 AU from the sun}. { In contrast, the same RT models fail to fit the UVS data inside the 10 AU boundary for both V1 \& V2 and beyond the outer 50 AU boundary for V1 \citep{que13}}. While a low level of the \lya\, solar flux presumably used by \citet{hal92} is propounded to explain the V2 new sensitivity, no further instrumental explanation is proposed for the enigmatic 243\% jump derived for the V1 UVS sensitivity \citep{que13}. In addition, we cannot find in Hall's 1992 thesis any link between the derived UVS \lya\, sensitivities and the solar flux level used by that author for his RT modeling, a problem that calls into question the jump in both the V1 and V2 calibrations.

First, it is important to stress that the V1 and V2 archive database also contains all the planetary airglow measurements obtained during the two spacecraft encounters with the outer planets. The enhanced new sensitivity of the UVS should strongly modify a planet's airglow emission levels, which calls for some caution before revising past studies. In addition, most of the outer solar system targets have been observed by several instruments simultaneously with V1 and V2 either during encounters or remotely \citep{she88,mcg92}. More generally, what is actually missing is a comprehensive approach that incorporates the sky background glow, the planetary diffuse emissions, and multi-missions' simultaneous observations of different targets in order to derive V1 and V2 calibration that is coherent with most existing data. For all these reasons, the large differences between the old and new calibrations of UVS call for a reassessment.

In the following, we undertake such a comprehensive approach, insisting on its consistency rather than on the model's sophistication. In Section 2, we compare Saturn airglow emission simultaneously observed by V1 and V2 (in situ) and IUE (remote) in order to check the consistency of any proposed UVS calibration. First, we use Saturn and sky-background high resolution \lya\, line profiles observed by the Goddard High resolution spectrometer (GHRS) onboard the Hubble Space Telescope (HST) in order to evaluate the imprint of the interplanetary hydrogen (IPH) absorption between Saturn and Earth. In that context, our goal is twofold: to assess the UVS new calibration, and to compare the consistency of our results with the IPH model used by \citet{que13} to derive their calibration. {In Section 3, } we model the He 58.4\,nm airglow of Jupiter based on the Voyager UVS with the goal of deriving the Jovian He abundance that we compare to the Galileo probe {\it in situ} abundance. In that way, we may obtain a further indication of the V1 and V2 sensitivity level based on distinct channels of the UVS detectors. {With the UVS original calibration confirmed, we discuss in section 4 the limitations of the global radiation transfer technique used in \citet{que13}, and propose a distinct approach based on the local RT technique, which proved to be efficient in several applications \citep{puy97,puy98,ben00}. In that context,  we recall how the inversion of the Voyager UVS sky-background \lya\ maps helped derive the heliospheric hydrogen local abundance at specific positions up to 40 AU from the Sun (section 4.2), and how Fermi glow interpretation of the sky-background \lya\ emission excess, a feature detected by UVS deep in the heliosphere, was at the origin of the discovery of the heliosphere distortion \citep{ben00}. Because the Fermi glow is controversial \citep{que06,que13}, we dedicate all of section 4.3 to addressing the issue, emphasizing the importance of the fact that despite their limited spectral resolutions, both Voyager UVS are stable enough to constraint key properties of the heliosphere without the need to modify their calibrations. In conclusion, we summarize all of our results and suggest few experimental solutions for efficient exploration of the heliosphere in the future.}

\section{Voyager UVS \lya\, channels calibration: reassessment}
{In the following, we use Saturn airglow emission simultaneously observed by Voyager and IUE to reassess the UVS calibration at \lya\ . However, to properly achieve the calibration diagnostic, one must first determine Saturn's \lya\, emission line, a key parameter that was missing in past studies. In that context, we first analyze Saturn and sky-background high resolution \lya\, line profiles observed by the Goddard High resolution spectrometer (GHRS) onboard the Hubble Space Telescope (HST) to determine that the line's width is large enough to show the imprint of the interplanetary hydrogen (IPH) absorption between Saturn and Earth (Section 2.2).  In the second step, we use the information on the Saturn \lya\, line width and a local inversion RT technique to directly compare UVS to IUE Saturn \lya\, brightnesses for the different UVS calibrations thus far proposed (Section 2.3).  We then show both the inadequacy of the newly proposed UVS calibration and the validity of the instrument original sensitivity derived after the Jupiter encounter. }

\subsection{Diffuse sources in the solar system: Cross-calibration of UV instruments}

Because the distance to the outer planets is large enough, the interplanetary medium that separates them from Earth is opaque enough in atomic hydrogen to leave an imprint on their \lya\, brightness. Therefore, comparing remote to {\it in situ} observations of the planet offers key constraints on both the IPH opacity at \lya\, and on the relative calibration between the observing instruments. {More generally, abundances derived from planetary airglow emissions could be validated only if the same species could be measured either in the atmosphere or by independent techniques, like solar and stellar occultations that are widely used on space missions \citep{bro81}}. In this respect, we consider here the H\,I abundance derived from the Saturn \lya\, airglow measured by both Voyager and IUE instruments, the He abundance in Jupiter derived from the planet's airglow measurements, and the Galileo probe measurement within the Jovian atmosphere.  For reference, the IUE calibration has been assessed many times during the lifetime of the spacecraft, yielding firm conclusions as to the sensitivity of the instrument and its time variation that should not exceed a few percent \citep{boh90}. {Because our discussion will mainly focus on the Voyager UVS calibration within the context of the Jupiter and Saturn airglow observations, all absolute fluxes used here are based on the standard calibration pipeline described in \citet{hol92}, taking into account the correction reported for the \lya\, channels} \citep{hal92}.

\begin{figure*}
\centering
\includegraphics[width=12.cm,angle=-90]{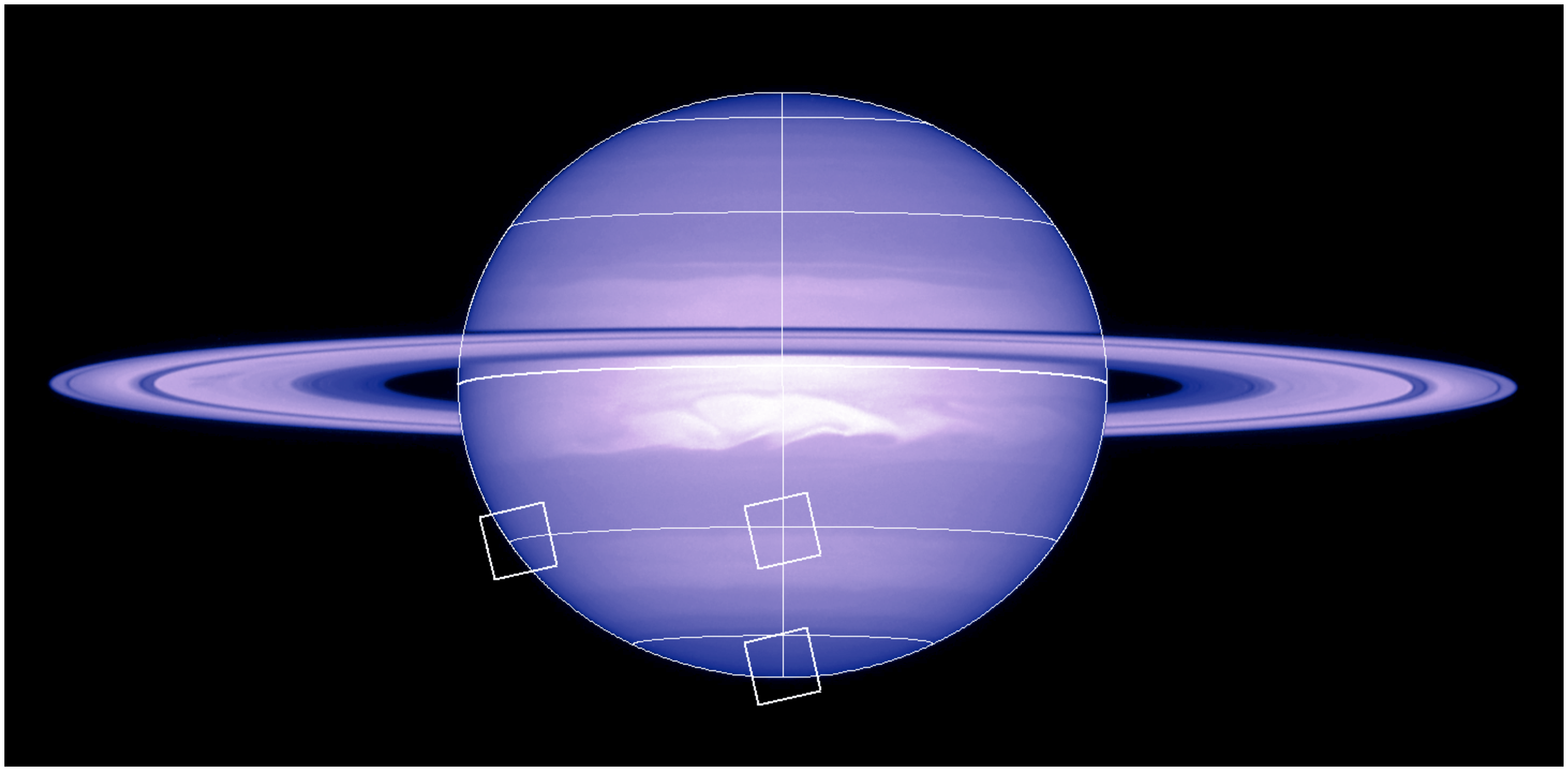}
\caption{Sketch of the GHRS slit positions on the Saturn disk shown for the December 24, 1996 observations (HST program GO 5757). The three exposures provide the disk averaged \lya\, spectrum used in this study. The GHRS large science slit is shown with the right angular size (1.7x1.7 arcsec$^2$). The image was obtained in the visible range with the Wide Field Planetary Camera (WFPC) very close in time to the GHRS \lya\, observations, and is shown here only for reference.} 
\end{figure*}

First, it is important to recall that IUE was used for remote monitoring of the Saturn \lya\, brightness over the period of 1980-1990. Such long-term strategy has many benefits, particularly in showing the strong correlation between the Saturn \lya\, airglow and the solar \lya\, flux, along with intrinsic temporal and spatial variations observed across the planetary disk \citep{mcg92,she09}. For instance, at the time of encounter of the Voyager 1 and 2 with Saturn respectively in 1980 and 1981, the \lya\, dayglow of the planet was observed nearly simultaneously by V1 UVS and IUE on day of the year (doy) 343 of 1980 and then by V2 UVS and IUE on doy 240 of 1981 (e.g., Table 1). {After the damage suffered by the V1 spacecraft during its crossing of the Jovian magnetosphere, cross-correlation of Voyager 1 UVS, Voyager 2 UVS, and IUE  observations of the Saturn \lya\, emission  has been used to assess the UVS calibration \citep{she84,she88,hal92}. The simultaneous measurements were also capitalized by \citet{puy97} and \citet{puy98}, leading to a self-consistent estimation of the IPH density}. Generally, the moving IPH absorbs the planetary emission line at a spectral range that depends on the orientation of the line-of-sight to the planet in the heliosphere with respect to the upwind direction ($\sim \pm 26\,$km/s). {However, in all those studies, a key ingredient was missing: the exact shape of the Saturn \lya\, line profile. Instead, it was approximated based on photometry or low-resolution observations and theoretical modeling of the excitation processes at the origin of the planetary emission line} \citep{mcg92,ben95,puy97}. Within that framework, it is important to accurately derive the shape of the planetary emission line to check whether the width is large enough to show noticeable extinction by the IPH gas for the spectral range of the Doppler-shifted absorption \citep{wuj79a}.

\subsection{HST observation of Saturn and sky-background \lya\, emission line profile}

To constrain the Saturn \lya\, emission line profile, we use the archive observations of the Saturn \lya\, dayglow obtained with the Goddard High Resolution Spectrograph (GHRS) onboard the HST using the Echelle grating and the large 1.74$\times$1.74 arcsec$^2$ aperture (HST program GO 5757). The HST program GO 5757 consisted of two distinct visits of the planet, one in October 1994 to obtain spatial scans across the planetary disk, and a second on December 24, 1996 but at fixed positions on the disk. Here, we decided to use the 1996 dataset because the corresponding Doppler separation between the Earth geocoronal emission line and the Saturn emission line was the largest ($\sim 30$\,km/s). In total, we use three distinct exposures for the planet. An additional exposure for the sky background was also obtained $\sim 2$ arcmin away from the planet (Figure 2), far enough from the HI cloud that surrounds Saturn \citep{she92}. For the planet, each exposure was obtained as a time sequence of 6 or 7 sub-exposures (for a total of roughly $\sim 900$\,s). Here, the three separate pointings thus far used correspond respectively to latitude 30$^{\circ}$ south near disk center (HST dataset z2jr1104t), latitude 30$^{\circ}$ south near sunlit limb (z2jr110et), and south pole (z2jr1109t) as sketched in Figure 1. 

\begin{figure}
\centering
\hspace*{-0.15in}
\includegraphics[width=9cm]{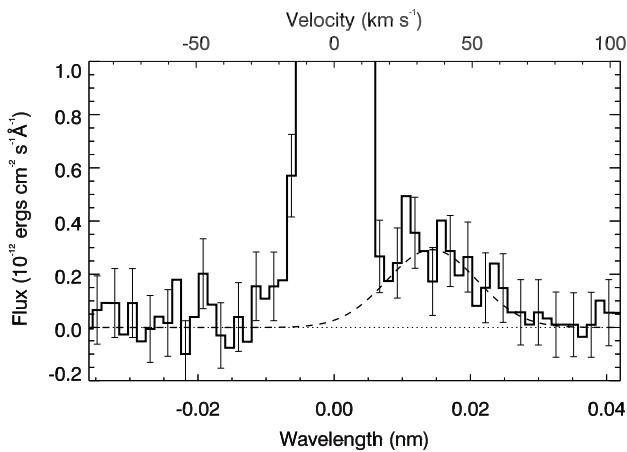}
\caption{ Sky-background \lya\, emission observed with HST/GHRS Echelle mode using the large science slit on December 24, 1996.  Spectra are binned by 4 to reduce the noise level and shown in the Earth reference frame. The line-of-sight was $\sim 2$ arcmin far outside the limb of Saturn. The IPH emission line toward infinity clearly appears on the red-wing of the strong Earth geocorona emission line. We assume Voigt profiles for the two lines that are convolved with the GHRS line spread function ($\sim 0.007$\,nm). Using the Levenberg-Marquardt least-squares technique, our best fit to the IPH line (dash) is $ 35.7\pm 2.0$\,km/s Doppler-shifted from the geocoronal line, and has a Voigt temperature of $ (2.7\pm1.8)\times 10^4$\,K and a brightness $ 520\pm 100$\,R. }
\end{figure}

{We compared the line profile obtained at the south pole location (allowing for the possibility of auroral contamination) to the latitude 30$^{\circ}$ south near disk center and to the average of three line profiles. Because the individual spectra are noisy, we cannot find any noticeable difference} (Figure 3). As documented by many observations of the Saturn aurora obtained with HST since 1994, a plausible explanation is that the south pole aurora could be very faint on December 24, 1996 (the observed system III long/lat of the slit were $\sim 108$$^{\circ}$/76$^{\circ}$S with a subsolar point latitude of 4$^{\circ}$S), in addition to the fact that the latitudinal extent of the Saturn aurora may not fill the whole GHRS slit \citep{ger05}. {Furthermore, for both the south pole and sunlit observations, the GHRS slit is partially filled by the emission from the planetary disk}, which may also explain the comparable level of signal obtained for the three pointings (see Figure 1). While the details of the individual line profiles are important for the study of Saturn's upper atmosphere, our primary goal here is to find that the line width is comparable for the three targets. 

\subsubsection{Geocoronal and sky-background contaminations subtraction}

{Most UV observations are scheduled when HST is in the Earth's shadow to reduce the geocoronal airglow contamination. Because HST observations start with a high contamination from the \lya\, geocoronal emission line at the Earth day-night terminator}, we dismissed the first exposure of each visit, which leads to a final $\sim 2700$\, s high-resolution line profile of the planetary emission and $\sim 900$\,s exposure of the sky-background obtained near the orbital position of Saturn. For the observed diffuse light sources, the GHRS slit is filled, which reduces the final resolution of the Echelle mode to $\sim 18000$. Here, it is important to stress that a wavelength lamp calibration observation was scheduled after each of the four HST visits, which helped calibrate the spectral line positions to better than $\sim 2$\,km/s (equivalent to $\sim 0.008$\, \AA) \citep{eme96}.

\begin{figure}
\centering
\hspace*{-0.15in}
\includegraphics[width=9cm,height=8cm]{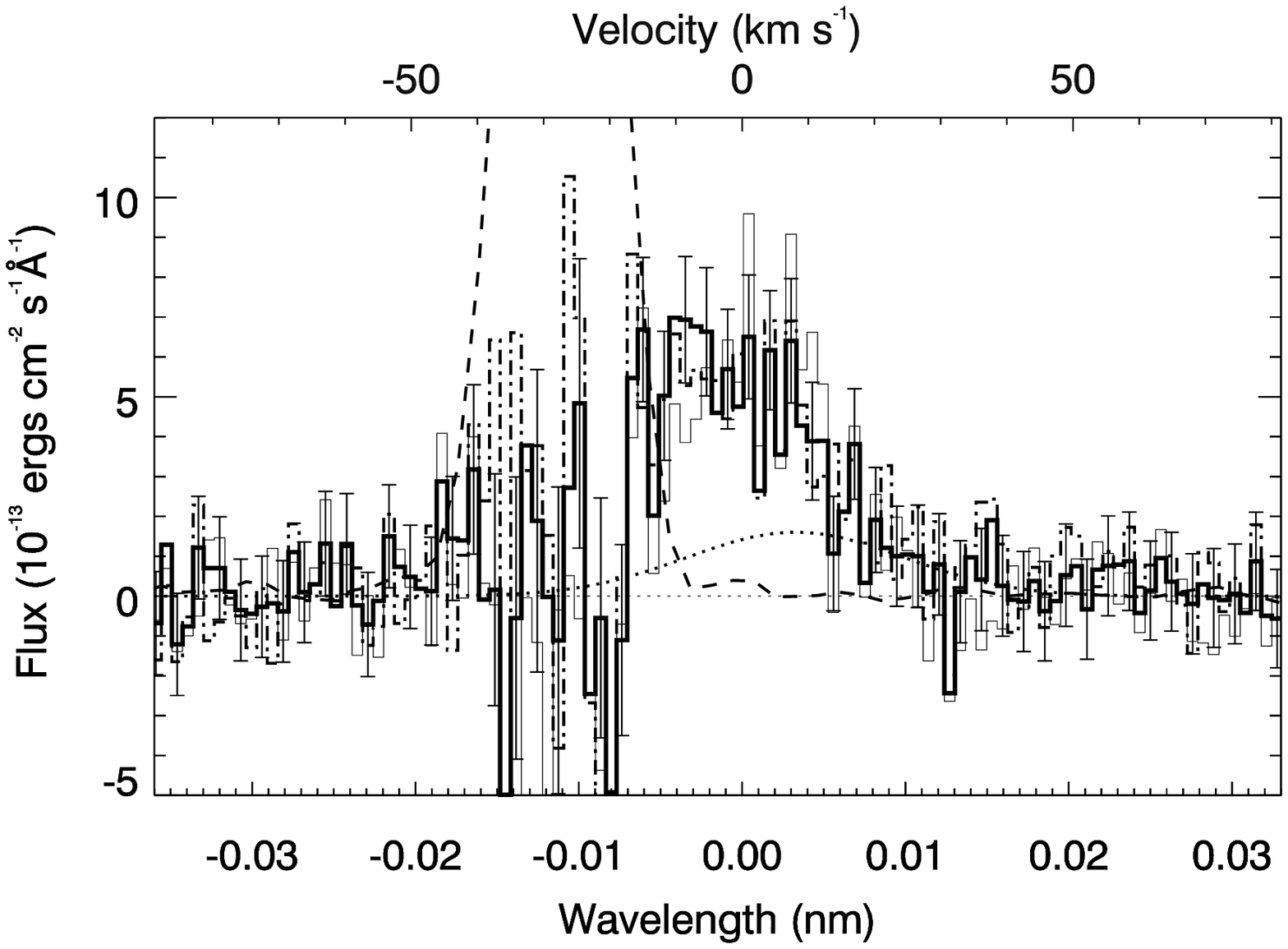}
\caption{ Saturn \lya\, emission line observed with HST/GHRS Echelle mode using the large science slit on December 24, 1996.  Spectra are binned by 2 to reduce the noise level. The Earth geocorona emission (dashed line) was observed $\sim 2$ arcmin  far outside the limb of Saturn $\sim 1.5$\, hours later, which corresponds to the following HST orbit.  The average IPH emission line toward Saturn (dotted) has the same FWHM as the line shown in Figure 2 but with a reduced brightness $ 280\pm100$\,R (compared to $520\pm100$\, R for the sky-background emission observed to infinity). The resulting spectrum (thick histogram, brightness $ 900\pm100$\,R) is the extracted Saturn emission line suffering an extinction by the IPH between the planet and Earth. For reference, we also show the emission line profiles obtained for the equatorial (solid thin line) and south pole (dash-dot line) targets, confirming the large width of the Saturn \lya\, emission line.}
\end{figure}

In the initial plan of the HST program GO 5757, observations were scheduled to use Earth's maximum orbital speed ($\sim 30.2$\,km/s), which allows for the largest Doppler shift ($ 35.7\pm 2.0$\,km/s) between the geocoronal contamination and the planetary lines. At the time of observation in December 1996, the line of sight pointing toward Saturn was $\sim 105^{\circ}$ away from upwind (around the crosswind region), which put the interplanetary emission line at an average Doppler shift of  $ 5.5\pm 2.0$\, km/s, as would be expected for that direction for an LIC speed of 26.4\, km/s \citep{wuj79b,kat11}. To achieve an accurate analysis of the spectral lines, a line spread function (LSF) of the instrument is required. For the GHRS in the Echelle and large slit mode, the LSF issue was carefully discussed by \citet{cla98} in the frame of the sky background observations and by \citet{rob98} in the general case.  With an accurate LSF on hand \citep{rob98}, the Doppler separation ($\sim 36$\, km/s) between the geocoronal and IPH lines allows for extraction of both lines, which later should help in properly subtracting the variable geocoronal emission from the planetary one (e.g., Figure 2). For that purpose, we modeled the faint IPH line as a Voigt function that we convolved with the GHRS Echelle LSF and subtracted from the sky background observation using the Levenberg-Marquardt least-squares technique \citep{vin11}. In that way, we obtain a geocoronal emission line in the same instrument conditions as for the planetary line (e.g., Figure 2). Having the Earth geocoronal and IPH lines separated makes it possible to scale each of them independently during the cleaning of the spectrum observed toward the disk of Saturn from its corresponding sky contamination.

\begin{table}[h]
\caption{ Best fit HST/GHRS Saturn \lya\, line profile: sensitivity to IPH models}
\begin{center}
\hspace*{-0.2in}
\scalebox{0.67}{
\begin{tabular}{cccc}
\hline

Model                              & model A & model B & Model C \\
LISM proton density ($cm^{-3}$) & 0.032  & 0.032 & 0.06                \\
Temperature primary (K)             & 6100   &6100 & 6020             \\
Temperature secondary (K)         & 16500 & 16500 & 16300             \\
Density primary at 90 AU ($cm^{-3}$)      & 0.088 & 0.058 & 0.035                \\
Density secondary at 90 AU ($cm^{-3}$)  & 0.088 & 0.058 & 0.06                  \\
IPH velocity primary at 90 AU  (Km/s)    & 28 & 28 & 28 					\\
IPH velocity secondary at 90 AU  (Km/s)    & 18 & 18 & 18 					\\
Earth-Saturn layer primary opacity 										&  0.361 & 0.24 & 0.14      \\
Earth-Saturn layer secondary opacity 										&  0.185 & 0.12 & 0.13      \\
Model line width (\AA) 								& 0.108 & 0.107 & 0.108       \\
Planet \lya\, brightness (R)                        & 1200  & 1000 & 970   \\
\hline
\end{tabular} }
\end{center}
\tablecomments{Best fit obtained for the HST/GHRS Saturn \lya\, line profile obtained after correction for the IPH absorption. { IPH model A is based on \citet{bzo08} but with a relatively higher density as reported in \citet{puy98}. Models B and C are based on \citet{bzo08}}.  It is interesting to note that the width of the derived Saturn \lya\, line profile is insensitive to the adopted IPH model. In contrast, after correction for the model IPH absorption, the integrated emission line is sensitive to the model adopted, yet with a brightness level, for all models assumed, that is consistent with IUE observations obtained around the precedent 1985-1987 solar minimum \citep{mcg92} }
\end{table}

It is now possible to properly clean the planetary emission spectrum first by scaling and subtracting the geocoronal emission line. However, it is important to {remark} that the intensity of the IPH emission line toward the disk of Saturn is different from the one measured toward infinity (far outside the planetary limb). In addition, for the sunlit limb and south pole pointing used here, the field of view may also contain a small contribution from the glow of the hydrogen cloud that surrounds the planet \citep{she92,she09}. As noted in early studies, removing the total IPH emission results in over-subtracting from the planetary emission the sky background emission that originates from behind the planet \citep{mcg92}. {Using the Levenberg-Marquardt least-squares technique, our best fit shows that for the observations toward the three targets (equatorial, sunlit, south pole) on the Saturn disk, the corresponding sky background emission line has $\sim$(0.3, 0.67, 0.67) times the brightness of the same line ($520\pm100$\,R) observed $\sim 2$\,arcmin away from the planet. The diagnostic was made on the red wing of the emission line of each of the individual targets, where the sky background contamination is the most prominent (e.g., Figure 3). {Independently of any attached statistical or systematic uncertainties, the sky-background emission toward Saturn is bound by two limiting values}: the brightness of the same line toward infinity as an upper limit, and a very weak emission at the noise level of the observation as a lower limit (e.g., no sky-background emission). In the following, we will also use the two limiting values of the sky-background emission to bind the width of the Saturn \lya\, emission line strongly, yet independently of the sky contamination observed and the related statistical uncertainties. 

\subsubsection{Sensitivity of the Saturn \lya\ line to the IPH absorption}
After the subtraction of the reduced IPH emission, the resulting emission line observed at the Earth orbits should correspond to the intrinsic planet's line that is partially absorbed by the IPH atoms that fill the space between the two positions \citep{mcg92,puy97}. {  The estimation of the IPH absorption is tricky because the gas distribution strongly depends on both the outer and inner boundary conditions assumed. For Saturn airglow correction, the problem was carefully discussed in \citet{mcg92}, yet the outer boundary effect was missing in their study. Because H I transport in the heliosphere is a complex problem, few approximations were proposed to obtain practical diagnostic tools that could be implemented for the inner heliosphere studies. For instance, the so called two-component, hot model was proposed as a helpful approximation to study the impact of neutrals on the H\, II pickup ions distribution measured in the inner heliosphere \citep{bzo08}. The model reflects the main properties of the heliosphere's neutral distribution, including the charge-exchange effect at the outer heliosphere \citep{bar93,bzo08,izm13}. In addition, the kinetic equation of neutral transport is solved separately for the primary and secondary populations, which correspond to the dominant species that compose the inner heliosphere hydrogen \citep{bar93}. At the outer boundary, which is postulated at the solar wind termination shock (TS) around $\sim 90$\, AU,  the velocity distribution is assumed to be Maxwellian for each of the two populations, which have properties that are related to the LISM bulk properties (HI density nH$_{-lism}$, temperature T$_{-lism}$, proton density np$_{-lism}$) via the Baranov-Malama self-consistent model \citep{bar93,bzo08,izm13}. 

The two-component hot model was applied to interpret H\,II pickup ions production rates in the inner heliosphere, which helped estimate the HI density level around the TS around $\sim 90$\, AU \citep{bzo08}. Recently, a more sophisticated and proper handling the 3D boundary conditions at the outer heliosphere used 3D MHD-Kinetic and radiation transfer simulations to help build the heliosphere H I distribution and the corresponding photons scattering, which have been useful to analyse sky-background reflected light measured by different spacecrafts  \citep{kat11,que13,fay13}.  { However, despite their sophistication, those models were not very successful in describing neither the inner heliosphere (inside 10 AU) or the outer heliosphere (beyond 50 AU) sky-background radiation observed, and they gave no further indication of the origin of the discrepancy found} \citep{que13}. A distinct approach used the invariance principle (optics) to derive the H I neutral density distribution by directly inverting sky-background reflected light maps measured by both Voyager 1 and 2 locally along their path up to 35 AU from the Sun \citep{puy97,puy98}. Interestingly, those authors were able to fit the inner heliosphere H I distribution using a Baranov-Malama kinetic model to derive a self consistent solution that relates the LISM intrinsic properties (H I density, ionization rate, etc.) to the inner heliosphere distribution, yet the solution is not unique \citep{puy98}.

In addition to missing the LISM magnetic field, which forces a pressure distortion on the whole heliosphere, most of the above-cited studies also miss the self-shielding effect of the solar radiation at Lyman-$\alpha\,$ by the IPH atoms that should strongly modify the radiation pressure force strength in the inner heliosphere, and consequently, the H I distribution in the inner heliosphere \citep{wuj79b}. {Recently, momentum transfer during photon scattering with H I atoms was properly estimated using Monte Carlo scattering model, leading to the important conclusion that the radiation pressure calculated fall-off with distance from the Sun may differ significantly from the ${1\over r^2}$ dependency typically assumed in the past \citep{fay15}}. Given the present conditions, it is difficult to come up with a self-consistent model for the IPH that would fit most sky-background radiation observations from the inner to the outer heliosphere \citep{ben15}. For those reasons, we consider  typical models thus far reported in the literature to describe the IPH in the 1-10 AU region, but we will not discuss the general problem of the IPH composition and distribution as those are beyond the scope of the calibration study conducted here. Therefore, in the inner boundary, we assume a total ionization rate $\beta\sim 7.5\times 10^{-7}\, {\rm s^{-1}}$ and a radiation pressure-to- gravity forces ratio that depends on the H I opacity radially (see section 2.3 for more detail). 

In order to test the sensitivity of our results to the IPH properties in the outer heliosphere, we consider three two-component models that are consistent with the Baranov-Malama model in the outer heliosphere boundary (A, B, and C) as defined in Table 1. Model A is based on the H I distribution derived for the inner heliosphere up to 35 AU by \citet{puy98}, a distribution that can be characterized by its relatively high local H I density at 10 AU. Model B and model C correspond to the two-component, hot models assumed in \citet{bzo08}. For model A, because the LISM proton density derived in \citet{puy98} ($\sim 0.04\, {\rm cm^{-3}}$) is close to model B, we assume the same model but with a larger IPH density. While the assumption might seem arbitrary, it is worthwhile because it shows the effect of a strong extinction by the IPH  \citep{puy98}.  All IPH models are calculated for a direction of ecliptic coordinates $\lambda\sim 1^{\circ}$ and $\zeta\sim -2.4^{\circ}$ of Saturn on December 24, 1996, which corresponds to a line of sight that is $\sim 105^{\circ}$ away from upwind. For that direction, the hydrogen layer between Saturn ($\sim 9.47$\,AU) and Earth ($\sim 0.99$\,AU) has a total opacity indicated in Table 1 for each model. 

Here, it is important to stress that the three models are considered for the calibration purposes only because no unique correction exists for the IPH extinction \citep{she84,mcg92}. Indeed, our purpose is not to derive the exact Saturn \lya\, line profile, but rather to verify that the width of the line is large enough to reflect the Doppler-shifted absorption produced by the H\,I atoms flowing between Saturn and Earth \citep[]{puy97}. In the first step, to estimate the sensitivity of the line width to the sky-background emission subtracted, we derive the best fit obtained for reference Model A, also showing the impact of subtracting the two limiting maximal and minimal values of the sky-background emission measured toward Saturn (Figure 4). In doing that, we test the sensitivity of the derived line width to a systematic error that would be much larger than the statistical uncertainty ($\sigma_B=100$\,R) attached to the HST/GHRS data. {For reference, it is helpful to recall that the minimal line width required to see a substantial IPH imprint is related to the Doppler shift that corresponds to the line of sight assumed. In the case of the HST 1996 observations, the minimal line width should be $\sim 5 \times10^{-3}$\,nm}. It is interesting to see that the width of the Saturn \lya\, line is bound between $ (9.2\pm 1)\times10^{-3}$\,nm (when subtracting 9/10$^{th}$, the maximum value measured to infinity, which corresponds to $\sim 4.7\times\sigma_B$) and $ (11.5\pm 1)\times10^{-3}$\,nm (1/10$^{th}$ {the maximum value measured to infinity}, which corresponds to $\sim 0.5\times\sigma_B$ )}. The derived Saturn \lya\, line width $(10\pm1)\times10^{-3}$\,nm is larger than the value previously used in \citet{puy97} and is, in all cases, large enough to bear the imprint of the IPH absorption. 

\begin{figure}
\centering
\hspace*{-0.15in}
\includegraphics[width=9cm]{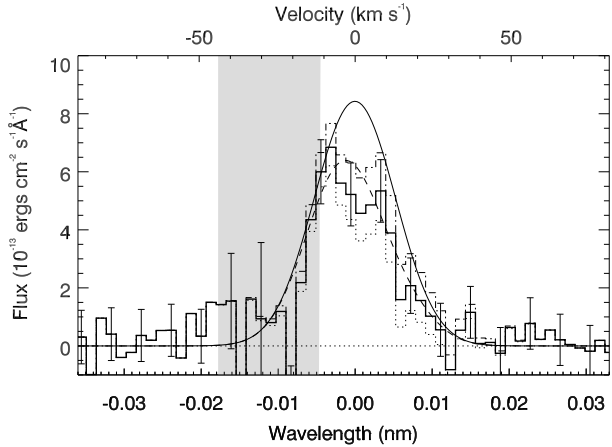}
\caption{ Saturn \lya\, emission line (histogram binned by 4) observed with HST/GHRS Echelle mode on December 24, 1996 and shown for three distinct corrections of the sky background emission toward the planet. The best fit (solid) is compared to two limiting corrections corresponding to the maximum (dots) and minimum (dot-dash) sky background emission subtracted. The shaded area corresponds to the spectral range of the geocoronal emission line contamination. We assume a gaussian profile for the planetary \lya\ line that is convolved with the GHRS line spread function ($\sim 0.007$\,nm). The IPH absorption by hot model A is applied and the resulting line profile (dash line) is compared to the observation (solid histogram). The Saturn intrinsic \lya\, emission line without any contamination (solid line) has a FWHM that is strictly bound between $0.009\pm0.001$\,nm and $ 0.011\pm0.001$\,nm corresponding to the two unrealistic extrema of the sky background emission subtracted (see text). Only the fit to the middle curve (solid) is shown for clarity. }
\end{figure}

In a second step, to test the sensitivity of our results to the IPH absorption assumed, we show in Figure 5 the best fit to the Saturn \lya\, line profiles before and after absorption by each of the three IPH models for the conditions of observation by HST/GHRS (12/1996). {First, for the weak opacity models B \& C, we remark that despite the large statistical uncertainties of the data, a slight difference exists between the planetary emission lines obtained before and after the IPH model absorption}. Second, it is interesting to note that the width of the line profiles derived is insensitive to the IPH model assumed ($\sim 0.01\pm 0.001$\,nm, e.g., Table 1). In contrast, the line brightness varies with the assumed model. As shown in Table 1, for the minimum of solar activity at which the HST/GHRS time of observations were scheduled, we obtain a planetary emission between $970\pm100$\,R and $1200\pm100$\,R for the three IPH models assumed, a brightness level that is consistent with IUE observations obtained around the precedent 1985-1987 solar minimum \citep{mcg92}. 

Our new analysis of the HST/GHRS observations of the Saturn \lya\, emission shows that the line profile is larger than previously assumed (e.g., Figures 4 $\& $ 5), a finding that confirms that the IPH atoms strongly absorb the planetary emission as predicted by previous studies at all possible Doppler shifts of the IPH line \citep{mcg92,puy97}. It is also interesting to note that the width of the planetary emission line is larger than predicted by resonant scattering of thermal hydrogen in the upper atmosphere of Saturn \citep{ben95}, a finding that requires further investigations to uncover any potential link with the hydrogen plume reported in Cassini/UVIS observations \citep{she09}, and more generally with the controversial problem of the heating of the thermosphere of Saturn \citep{smi07,she09,odo14}. Interestingly, the line broadening observed for the Saturn \lya\, line brings to mind a similar process observed by IUE and HST/GHRS in the bulge region in the upper atmosphere of Jupiter \citep{ben93,eme96}. It is outside the scope of this paper to evaluate the impact of the new HST/GHRS line profile reported here on the thermal structure of Saturn, a study that we leave for the near future.

\subsection{Saturn \lya\, emission: A lever to assess Voyager UV spectrometers calibration}
Given the important new result regarding the width of  the Saturn \lya\, emission line, which is large enough to bear the absorption by the moving IPH embedded between the planet and the Earth orbit, we can proceed using that line profile as a lever to test for the different Voyager UVS calibrations. In Table 2, we list the planetary \lya\, emission brightness simultaneously observed by IUE (remotely) and Voyager V1 and V2 (in situ), respectively, in November 1980 and August 1981, both for the traditional UVS calibration \citep{hal92,hol92} (hereafter VOY92) and the new calibration proposed by \citep{que13} (hereafter ISSI13). 

\begin{figure}
\centering
\hspace*{-0.15in}
\includegraphics[width=8.cm]{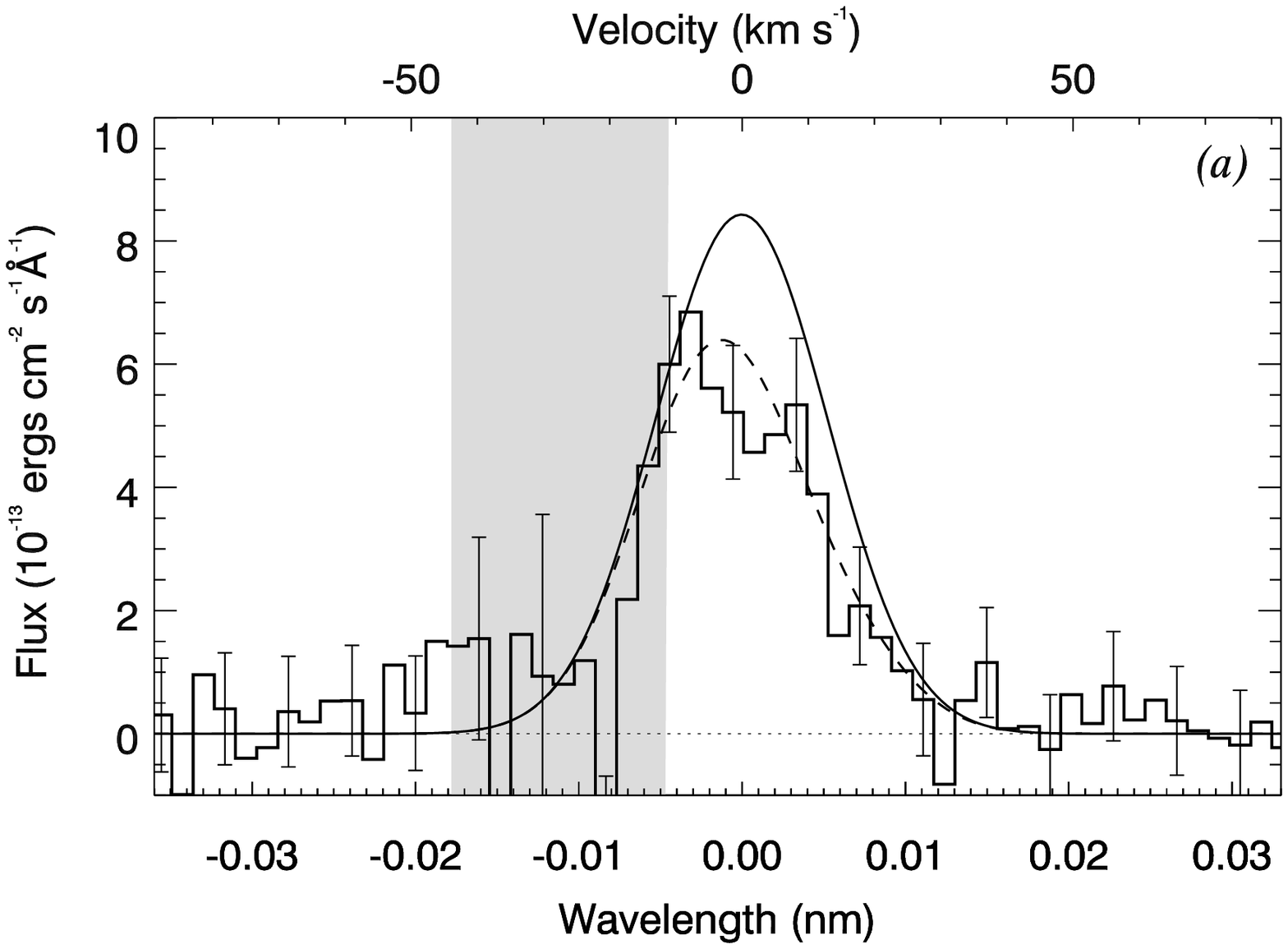}
\includegraphics[width=8.cm]{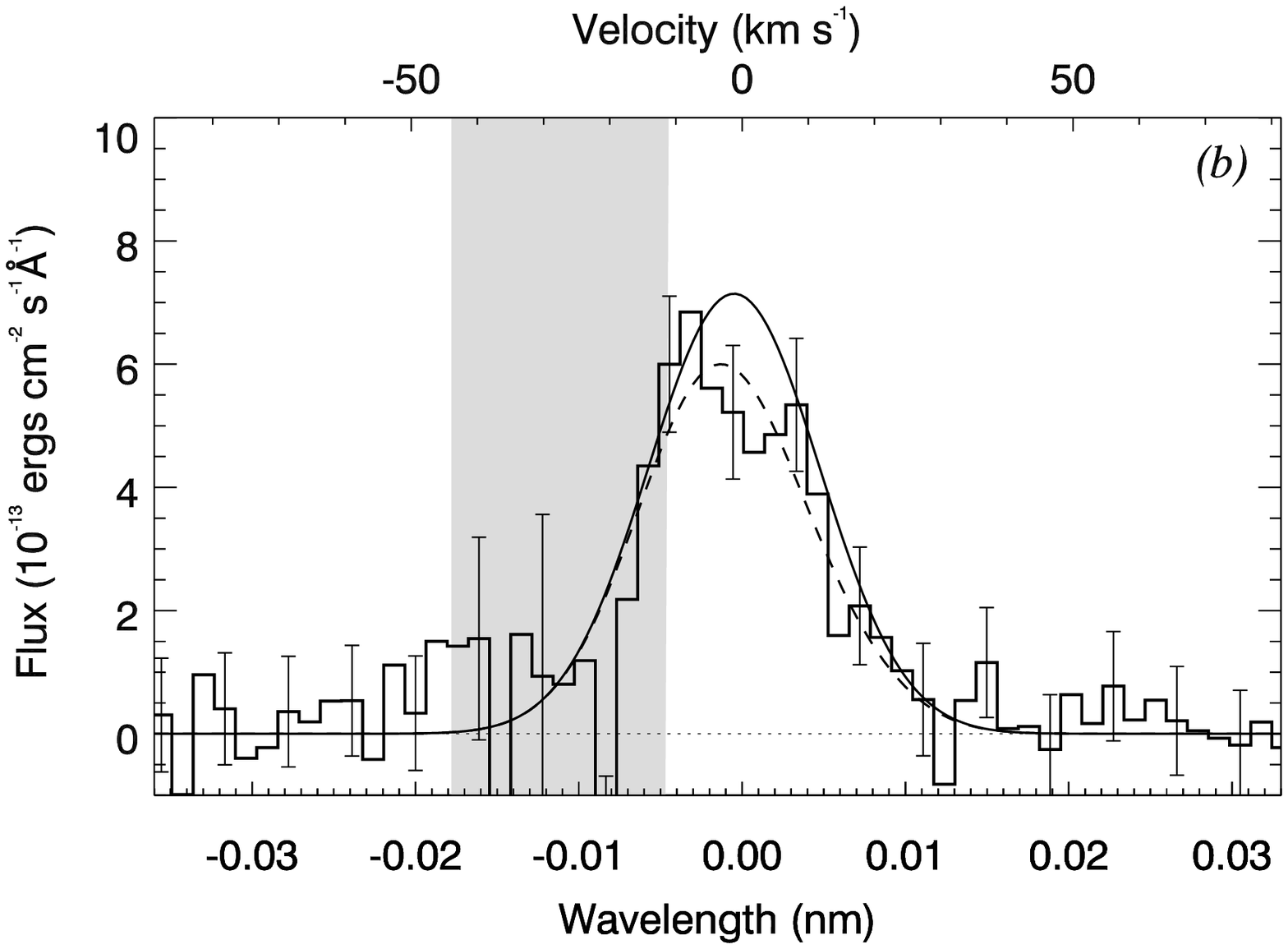}
\includegraphics[width=8.cm]{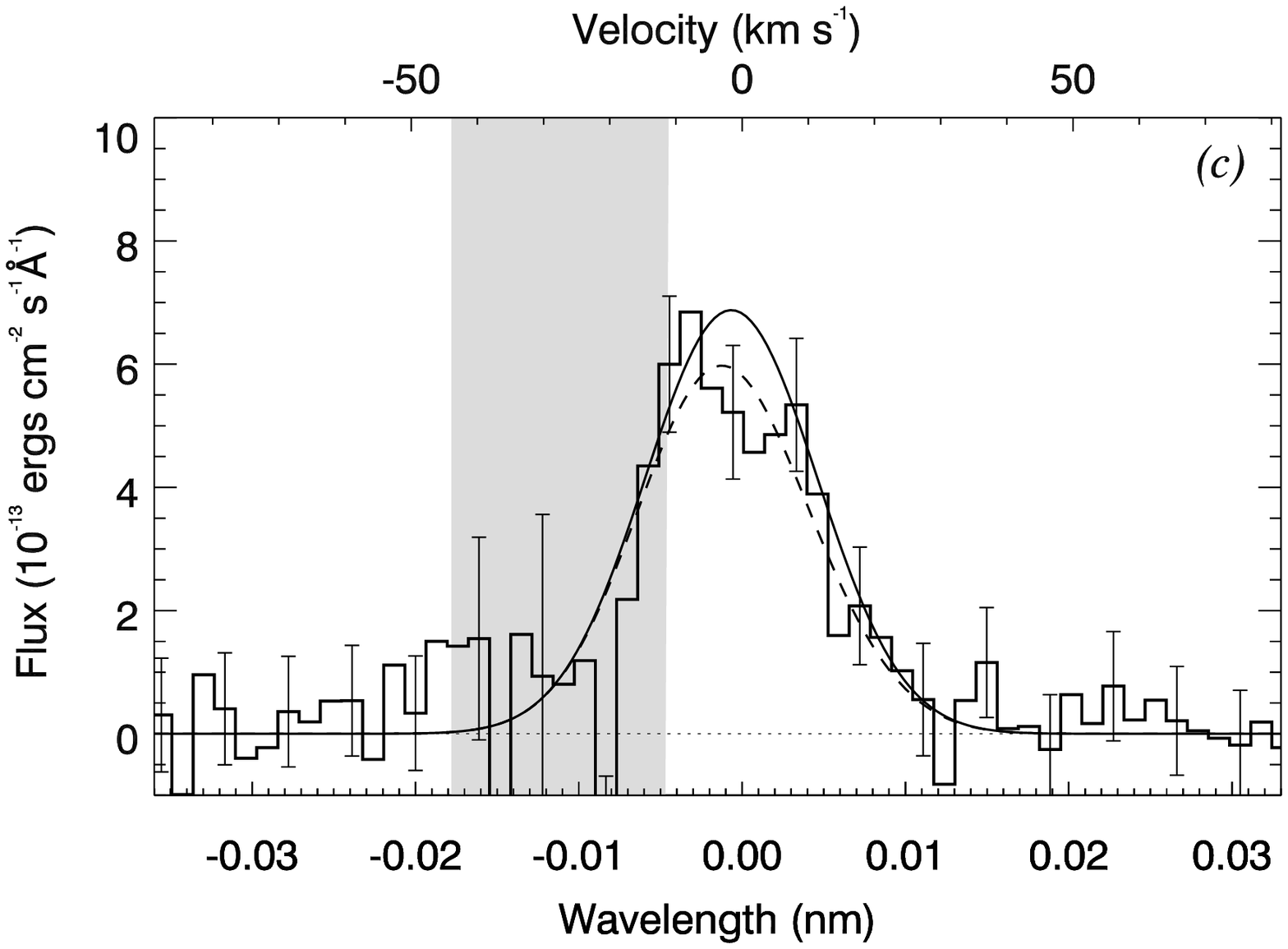}
\caption{Saturn \lya\, emission line (binned by 4) observed with HST/GHRS Echelle mode on December 24, 1996. The shaded area corresponds to the spectral range of the geocoronal emission line contamination. { We assume a gaussian profile for the planetary \lya\ line that is convolved with the GHRS line spread function ($\sim 0.007$\,nm). The IPH absorption (two-component, hot model) is applied and the resulting line profile is compared to the observation (dash line). The resulting Saturn intrinsic \lya\, emission line without any contamination (solid) has a FWHM$\sim 0.01\,$nm for all IPH models defined in Table 1. We use the Levenberg-Marquardt technique to solve the least-squares problem \citep{vin11}. {\it (a)} Model A . {\it (b)} Model B. {\it (c)} Model C. All derived brightnesses are comparable to IUE observations made during the 1985-1987 minimum solar activity (e.g., Table 1)} \citep{mcg92}. }
\end{figure}

{ In the following, we use the same invariance principle inversion technique as described in \citet{puy97} and \citet{puy98} to derive the IPH number density required in order to match IUE and Voyager UVS observations of the Saturn \lya\, brightness. First, it is important to stress that in deriving their results for the inner heliosphere (1-10 AU), \citet{puy97} assumed a parametric hot model with a temperature T$_{\infty}$=8000 K and V$_{\infty}$=20 km/s for the IPH layer separating Earth (IUE) from Saturn (V1 \& V2). However, as described in the previous section, sophisticated kinetic modeling of the heliospheric hydrogen clearly show that a two-component hot model is more suitable to capturing the impact of the charge-exchange process that operates near the heliopause to produce, at minimum, the secondary hot H I population \citep{bar93,puy98,bzo08,izm13}. In that context, the question is how the relative weight of the secondary population density (compared to the primary population) affects the boundary conditions at the 10 AU position and the results thus far reported in \citet{puy97}, particularly the opacity between Earth and Saturn. The issue was noted yet not fully explored in \citet{puy98}. In addition, most studies of the IPH distribution near the sun neglected the effect of self-shielding by atomic H, a process that may strongly absorb the solar \lya\, line and consequently modify the radiation pressure force strength versus distance along the Earth-Saturn line-of-sight  \citep{wuj79a}}. For reference, \citet{puy97} also derived in a self-consistent way the ionization rate by solar wind particles, showing that the parameter is well described by theoretical models related to the daily measurements of the solar wind parameters \citep{bzo13}. Finally, it is important to stress that \citet{puy97} and \citet{puy98} discussed the sensitivity of their results to the Voyager UVS calibration but only within the attached 20\% uncertainty known at that time. With that in mind, it would help to know how sensitive the inversion technique and the derived IPH density to all the uncertainties listed above.

At this point, it is essential to recall that the IPH models used here are parametric models intended only to fit the opacity produced by the density distribution embedded between Earth and Saturn.  First, as described in the previous section, the two-component model is a suitable approximation for describing the nature of the inner heliosphere's atomic hydrogen that is mainly composed of two populations-- namely, the primary interstellar medium species, and the secondary species that suffered a first charge-exchange with the plasma background in the outer heliosheath (e.g. \citet{bar93,izm13,zan15}). Second, in considering three IPH models with distinct LISM properties, we test the sensitivity of our results to the outer heliosphere's boundary condition. To ease the implementation of the fitting process, we make use of three existing models thus far published in the literature: the two-component model 1 \& 2 defined in \citep{bzo08}, and the two-component model defined in \citet{izm13}. All parameters of the three models are defined in Table 3.  { For each of these models, considered separately,} we operate the fitting process by keeping the velocity, temperature, and the ratio of the primary to the secondary population densities unchanged, but varying the primary species density versus the ionization rate $\beta$ in order to match left-hand to right-hand of Eq. 4 of \citet{puy97} :

\begin{equation}
I_{IUE}^{Sat} - I_{IUE}^{Sky} = T \cdot I_{UVS}^{Sat} - T\cdot I_{UVS}^{Sky} - I_{screen}
\end{equation}
where I$_{screen}$ is the sky-background emission blocked by the disk of Saturn and $T$ is the transmission function that depends on the photon frequency and on the opacity of the Earth-Saturn layer \citep[for more details, see][]{puy97}}. { We repeat the same fitting process for each of the three distinct models (1-3) in order to test the sensitivity of our approach to different ratios of the primary to secondary population densities}.

 {Because matching the brightness from each the IUE observations to that of each respective Voyager UVS depends only on the Earth-Saturn layer opacity}, we determined that we would display our fitting results using the pair ( $\beta$, $n_{10 AU}$), where $n_{10 AU}$ is the total hydrogen density at 10 AU. { As we show below, our results are insensitive to the assumptions made on the outer heliosphere conditions, particularly to the relative weight of the secondary to primary populations (or equivalenty, the LISM proton density) assumed for each of the three main models considered}--results that, in turn, make our conclusion about the UVS calibration free of any modeling. The extrapolation of the derived densities at 10 AU to the outer heliosphere region is beyond the scope of the present study. 

To evaluate the radiation pressure ratio to gravity at any radial distance $r$ from the Sun, we use $\mu(r)=3.473 \times F_{\odot} \exp^{-\tau(r)} - 0.412$, where $\tau(r)$ is the opacity between the Sun and the position r \citep{bzo13,eme05}. If we include the VOY92 and ISSI13 calibrations, we have a total of twelve cases for matching remote (Earth) to {\it in situ} (Saturn) \lya\, observations: namely, IUE \& V1 (Model 1), IUE \& V1 (Model 2), IUE \& V1 (Model 3), IUE \& V2 (Model 1), IUE \& V2 (Model 2), and IUE \& V2 (Model 3) for either VOY92 or ISSI13 calibration. For each of these twelve cases, the exercise consists in finding the best density level at 10 AU for a set of ionization rate values ($\beta$\,) that allows matching the IUE to the Voyager UVS \lya\, brightnesses. As IPH models here are parametric models used only to describe the IPH distribution inside the Earth-Saturn layer, the same fitting process thus entails finding the ($\beta$, $n_{10 AU}$) relation for each of twelve cases, keeping in mind that each of the three IPH models considered are related to the outer heliosphere's physical parameters via the ratio $n_{sec}/n_{prim}$ assumed at 90 AU and the two components' temperatures and bulk velocities  (e.g., Table 3).

\begin{table*}
\caption{ Saturn and sky-background \lya\, observations}
\begin{center}
\hspace*{-0.1in}
\scalebox{0.6}{
\begin{tabular}{ccccccccc}
\hline
UV instrument & Observation date &  I$_P$  (R) $^a$ & I$_{Sky}^{\infty}$ (R)$^b$ &  (I$_P$ - I$_{Sky}^{\infty}$) (R)& Sun-Earth-Saturn & Observer-planet  & Solar flux$^d$  & Calibration method \\
 &  &  & & &angle$^c$ &  Distance (AU) & & \\
V1 UVS  & 12-11-1980 & 3300$\pm 200$  & 950  & 2350 &44$^{\circ}$L  &0.001&5.2 &VOY92\\
V1 UVS  & 12-11-1980 & 1360$\pm 80$  & 390  & 970   &44$^{\circ}$L &0.001   &5.2 &ISSI13\\
V2 UVS  & 26-8-1981  & 3000$\pm 200$   &1000 & 2000 &35$^{\circ}$T  &0.001 & 5.8 &VOY92\\
V2 UVS  & 26-8-1981  & 1920$\pm 130$   & 640  & 1280 &35$^{\circ}$T  &0.001 & 5.8&ISSI13\\
IUE        & 17-11-1980 & -       & -         & $ 1200\pm 100$  &48$^{\circ}$L  &9.51  & 5.3 &BOH90$^e$ \\
IUE        & 10-8-1981   & -       & -         & $ 1100\pm 100$  &49$^{\circ}$T  &9.59   &5.6&BOH90\\
\hline
\end{tabular} }
\end{center}
\tablecomments{ { (a)  I$_P$ = (I$_{Sat}$ + I$_{Sky}^{Sat}$) where I$_{Sat}$ is the intrinsic emission from the planet, and I$_{Sky}^{Sat}$ is the sky background emission toward Saturn with the planetary disk occulting the region behind. For Voyager in situ observations, I$_{Sky}^{Sat}$ vanishes \citep[for more details, see][]{puy97}. (b) I$_{Sky}^{\infty}$ is the sky background emission measured toward infinity by the quoted instrument at the indicated date and position. { (c) This angle allows to estimate the right solar flux taking into account the $\sim 25$ days rotation period of the Sun. L is for Saturn leading the Sun and T is for Saturn trailing the Sun (see JPL Horizon ephemerides for more details). (d) Unit is $10^{11}$ photons/cm$^2$/s}. (e) The IUE calibration used by \citet{mcg92} is the most updated sensitivity reported in \citep{boh90}. Note that for IUE, the sky background emission is derived from the same observation using a portion of the slit outside the planetary disk, which highlights that only the difference between the planetary and sky emissions is reported (e.g., see \citet{mcg92} for more details).}}
\end{table*}

In order to properly achieve the comparison between IUE and UVS observations, we should also evaluate the solar flux and the sky background emission level around Saturn for each date of observation. For the solar flux, we use the solar irradiance database provided by the Colorado group at http://lasp.colorado.edu/lisird/lya/. Using the JPL Horizon ephemerides, we could derive the right level of the solar flux that corresponds to each date of observation taking into account the solar rotation effect for the hemisphere that faces the planet (see Table 2). For the sky background emission around Saturn, we used V1 and V2 UVS scans obtained during the two spacecraft encounters with the planet \citep{ben95,yel86}. { By matching right hand to left hand of Eq. 1, we first attempt  to evaluate the sensitivity of the derived IPH density to the assumed parametric Model 1 to Model 3. As shown in Figure 6 (top curves), the best fit density distributions obtained for the cases  IUE \& V1 (model 1), IUE \& V1 (model 2), and IUE \& V1 (model 3) are comparable for the VOY92 calibration. We also obtain a similar conclusion using IUE \& V2 (model 1), IUE \& V2 (model 2), and IUE \& V2 (model 3) for the ISSI13 calibration (e.g., curves at bottom of Figure 6). Obviously, our best fits are not sensitive to the assumed parametric IPH model, a result that is not surprising because the difference between the IUE and Voyager UVS brightnesses is related to the IPH absorption integrated over the whole line that is mainly sensitive to the amount of H I atoms present between Earth and Saturn, a quantity that is governed by the total density of all absorbing species and the ionization rate.}

In the following, we discuss our results using model 1 (or equivalently model 2 or model 3) as a reference. Using Table 2, it is interesting to note that for the new ISSI13 calibration (correction by a factor 243\%), the V1 encounter Saturn brightness drops below the value measured remotely by IUE for the same period. {As shown in section 2.2.2, for most existing IPH models, the Saturn-Earth layer is opaque enough to reduce the brightness level measured at Earth compared to the level measured at the planet's position. In that context, one would expect a lower brightness level at Earth, a result that contradicts the ISSI13 brightness levels shown in Table 2}. Even in the unrealistic, extreme condition of empty space between Saturn and Earth with no absorption by the IPH atoms, the brightness level observed by IUE is not compatible with the ISSI13 calibration unless we modify the IUE calibration itself. The latter assumption is unrealistic because it would require a strong jump (243\% to 156\%) in the IUE sensitivity between 1980 and 1981 in order to be marginally consistent respectively with V1 and V2 observations of Saturn. Furthermore, the IUE calibration has been assessed many times in the past with changes that never exceed a few percent \citep{boh90}. The indisputable facts above prove that the new sensitivity enhancement of the \lya\, channels proposed by \citep{que13} for V1 UVS is absolutely unacceptable.

\begin{figure}[]
\centering
\hspace*{-0.15in}
\includegraphics[width=9cm]{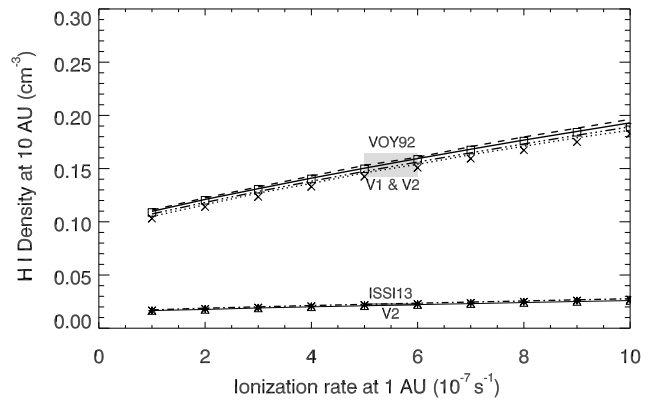}
\caption{{ Best fit for the pairs ($\beta$, n$_{10 AU}$) of the parametric two-component hot models using Eq. 1 to match IUE (1 AU) to the V1 \& V2 ($\sim 9.5$\,AU) observations of the \lya\, emission of Saturn, respectively in 1980 \& 1981. VOY92 refers to the Voyager reference calibration and ISSI13 corresponds to the new calibration recently proposed by \citep{que13}. For the VOY92 case we show the following curves  dash-3dots: (V1, model 1), solid: (V1, model 2), square: (V1, model 3), 
X: (V2, model 1),  dash: (V2, model 2), dot: (V2, model 3). For the ISSI13 V2 calibration, we show  triangle: (V2, model 1),  dash-dot: (V2, model 2), star: (V2, model 2). For the VOY92 calibration case, the best solution is obtained with an IPH density at the position of Saturn $n_{10 AU} = 0.15\pm 0.02\, cm^{-3}$ (for an angle of $\sim 73^{\circ}$ from upwind). For the ISSI13 V2 calibration, a marginal solution exists with a density at 10 AU $n_{10 AU} = 0.022\pm 0.004\, cm^{-3}$, at the limit of the noise level. In contrast,  no solution exists for the ISSI13 V1 calibration. The shaded areas shows the range ($5-6\times 10^{-7} {\rm s^{-1}}$) of the ionization rate at 1 AU $\beta$\, that best corresponds to the period of observation \citep{bzo13}.}}
\end{figure}

For the V2 ISSI13 calibration (correction by a factor 156\%), the comparison of the IUE to V2 \lya\, observations of Saturn leads to a solution that is shown in Figure 6 (triangle (Model 1), dash-dot (Model 2), and star (Model 3)). For the same ionization rate in the range $5-6\times 10^{-7} s^{-1}$, the best two-component, hot model solution has an IPH local density  $n_{10 AU}= 0.022\pm 0.004\, {\rm cm^{-3}}$ in the region around $ 10$\, AU at an angle $\sim 64^{\circ}$ from upwind. This solution is marginal because the difference between the corrected V2 \lya\, brightness ($\sim 1175$\,R corresponding  to the right-hand side of Eq. 1) and the IUE value ($\sim 1100$\, R) is at the noise level of the measurements. The corresponding IPH density is very weak as if the Earth-Saturn layer is mostly empty. { In addition, despite the similar conditions of solar wind and radiation parameters that prevail for the 1980-1981 period (less than 10\% variation), the non-existence of solution for the IUE/V1 observations in 1980 when using the ISSI13 calibration  casts doubt on the derived solution for 1981}.

In contrast, for the V1 \& V2 reference calibration (VOY92), the comparison of IUE to both V1 \& V2 \lya\, observations of Saturn leads to a solution for the IPH density that is depicted in Figure 6. For the ionization rate in the range $ 5-6\times 10^{-7} s^{-1}$ that prevails around 1980-1981 period \citep{bzo13}, the best model solution has  an IPH local density  $n_{10 AU}= 0.15\pm 0.02\, {\rm cm^{-3}}$ in the region around $\sim 10$\, AU at an angle $\sim 73^{\circ}$ from upwind, which is consistent with results previously obtained by \citet{puy98}. For reference, we obtained similar results (not shown) when using hot models with the appropriate boundary conditions as proposed in \citet{izm13}. Matching IUE to UVS observations leads to a total HI density at 10 AU that is independent of any assumption made on the outer heliosphere boundary.  All our results are summarized in Table 3. 

Here, it is important to stress that the degeneracy noted in the derived IPH density at 10 AU for the different IPH models used with distinct LISM ionization rate makes its extrapolation to the outer regions difficult because the exact weight of the primary to the secondary populations remains unconstrained. Other observational constraints should be used in the future in order to accurately derive the external LISM hydrogen density from inner heliosphere observations. For reference, using the Baranov-Malama model, \citet{puy98} could fit the HI distribution measured in the inner heliosphere using a LISM H I density $nH_{-lism}\sim0.24\pm 0.05\,{\rm cm^{-3}}$ for a LISM proton density of $np_{lism}\sim0.043\,{\rm cm^{-3}}$. {The solution is not unique, yet it is consistent with high HI density required in the outer heliosphere from independent 3D MHD-Kinetic and RT modeling of UVS observations \citep{kat16}}.

In conclusion, using the results of \citep{puy97,puy98} and the re-assessment provided above, we can see that the flux levels observed by IUE and V1 and V2 when using the VOY92 calibration are consistent with each other and with the IPH model that is assumed locally in the earth-Saturn layer. With the uncertainty of 30\% on the different parameters, our solution and those reported in \citep{puy97,puy98} are also consistent with the IPH model used in \citep{que13}. All of these simple facts show that the new sensitivity enhancement of the \lya\, channels proposed by \citep{que13} for V1 and V2 UVS is definitely unacceptable.

\begin{table*}
\caption{Matching IUE and Voyager UVS Saturn \lya\, observations: IPH model solutions}
\begin{center}
\hspace*{-0.1in}
\scalebox{0.6}{
\begin{tabular}{cccccccccc}
\hline
Model &   $n_{sec}/n_{prim}$ & T$_{prim}$ & T$_{sec}$ & V$_{prim}$ (km/s) & V$_{sec}$ (km/s) & Earth-Saturn layer & Earth-Saturn layer &  I$_{screen}$$^a$ (R)  & Total H I density$^b$    \\
& (at 90 AU)  &  & & &  & opacity (primary) & opacity (secondary) & & n$_{10 AU}$ (cm$^{-3})$ \\
IUE \& V1 (model 1, VOY92)  & 1.73 & 6020 & 16300 & 28.5 & 18.7 & 0.68 & 0.61 & 348 &  0.152\\
IUE \& V1 (model 2, VOY92)  & 1.00 & 6100 & 16500 & 28.2 & 18.5 & 0.88 & 0.45 & 340 &  0.155\\
IUE \& V1 (model 3, VOY92)  & 1.45 & 6840 & 18126 & 28.0 & 16.6 & 0.73 & 0.54 & 342 &  0.154\\
IUE \& V2 (model 1, VOY92)  & 1.73 & 6020 & 16300 & 28.5 & 18.7 & 0.64 & 0.56 & 351 &  0.147\\
IUE \& V2 (model 2, VOY92)  & 1.00 & 6100 & 16500 & 28.2 & 18.5 & 0.92 & 0.47 & 356 &  0.156\\
IUE \& V2 (model 3, VOY92)  & 1.45 & 6840 & 18126 & 28.0 & 16.6 & 0.72 & 0.54 & 355 &  0.150\\
IUE \& V1 (model 1, ISSI13)  & 1.73 & 6020 & 16300 & 28.5 & 18.7 & - & - & - & -\\
IUE \& V1 (model 2, ISSI13)  & 1.00 & 6100 & 16500 & 28.2 & 18.5 &  - & - & - & -\\
IUE \& V1 (model 3, ISSI13)  & 1.45 & 6840 & 18126 & 28.0 & 16.6 & - & - & - & -\\
IUE \& V2 (model 1, ISSI13)  & 1.73 & 6020 & 16300 & 28.5 & 18.7 & 6.8$\times 10^{-2}$ & 8.2$\times 10^{-2}$ & 197 & 0.022\\
IUE \& V2 (model 2, ISSI13)  & 1.00 & 6100 & 16500 & 28.2 & 18.5 & 0.1 & 6.7$\times 10^{-2}$ & 199 & 0.023\\
IUE \& V2 (model 3, ISSI13)  & 1.45 & 6840 & 18126 & 28.0 & 16.6 & 7.9$\times 10^{-2}$ & 7.7$\times 10^{-2}$& 198 & 0.022\\

\hline
\end{tabular} }
\end{center}
\tablecomments{ { Derived IPH opacity (for each of the primary and secondary populations), and corresponding total H I density required for matching IUE to V1 \& IUE to V2 Saturn \lya\ brightness using Eq. 1 for the different cases considered: VOY92 and ISSI13 calibrations and three distinct parametric two-component models (defined by the primary population temperature T$_{prim}$ and velocity V$_{prim}$, the secondary population temperature T$_{sec}$ and velocity V$_{sec}$, and the ratio of the secondary to the primary component densities $n_{sec}/n_{prim}$). The derived IPH total density (last column) corresponds to the sum of the primary and secondary population densities evaluated as a mean value over the range $\beta=5-6\times 10^{-7} {\rm s^{-1}}$ as shown in Figure 6. Our study shows that the total H I density is not sensitive to using Model 1, Model 2 or Model 3, which is consistent with the local inversion technique used}. For the ISSI13 calibration (243\%), it is impossible to match IUE and V1 Saturn \lya\, measurements. For V2, a marginal solution exists at the limit of the noise level. For the standard VOY92 calibration, a common solution exists for both V1 \& V2 independently of the IPH model assumed. (a) I$_{}screen$ is the sky-backround \lya\, emission that is blocked by the disk of Saturn and over-subtracted in the IUE measurement 
I$_P$ - I$_{Sky}^{\infty}$ \citep{puy97}. (b)  Total H I density is the best fit (n$_{prim}$ + n$_{sec}$) at 10 AU.}
\end{table*}

\section{Jupiter airglow: UVS He 58.4\,nm channels calibration}

We follow the same paradigm using airglow from planetary atmospheres to infer their main species abundance. The technique was recently applied to Jupiter and Saturn He abundance using V1 and V2 UVS observations of He 58.4\,nm dayglow made respectively in 1980 and 1981 \citep{ben15}. Here, we implement the technique as a lever to test the Voyager UVS calibration at the He 58.4\,nm channels using the He 58.4\,nm dayglow observed for Jupiter by Voyager 1 ($6.5\pm 1.1$\,R) and Voyager 2 ($5.7\pm 1.2$\,R) \citep{ver95,ben15}. For that purpose, we compare the Jovian He abundance derived from the Voyager UVS airglow to the one measured {\it in situ} by the Galileo probe  \citep{von95}. As shown in Table 3 of \citet{ben15}, the He mixing ratio $He /H_2 = 0.16 \pm 0.03$ derived from Voyager UVS nicely matches with the Galileo probe measurement of $He /H_2 = 0.156 \pm 0.006$ \citep{von95}. The coincidence between the two results confirms that the V1 and V2 UVS calibration used is correct \citep{hol91}. The uncertainty that appears in the derived He abundances shows that a 30\% uncertainty on the UVS sensitivities is also possible at He 58.4\,nm channels. To summarize, for the very distinct \lya\, and H 58.4\,nm channels, the Voyager UVS calibrations do not require any substantial revision beyond the attached 30\%. 

\section{Why the sky background modeling does not work for calibration: A few hints}
With the conclusions derived above, a legitimate question immediately presents itself: why is it that sophisticated models of sky background \lya\, emission do not provide the correct calibration?

\subsection{Global radiation transfer approach: an open issue}
For instance, the sky background is the target most observed by several space missions over the last four decades. Because Pioneer and Voyager UV instruments cover extended regions of the heliosphere over large time spans, the two databases have been extensively used, yet with a constant failure to correctly reproduce the brightness level over time and space unless the instrument calibration is revised up or down \citep{gan06,que13}. Here, it is crucial to realize that the problem is quite complex because it addresses a very extended, optically thick, moving medium illuminated by a time-variable radiation source, the Sun. Generally, the entire medium is included with appropriate boundary conditions, and the scattering is described either by using the Monte Carlo method or by making approximations of the number of photons diffusion \citep{kel81,hal92,sch99,que00,pry08,kat11,fay13}. Hereafter, we call the modeling concept described above as the global RT technique.  In most applications of the global RT technique, kinetic description of H\,I and 2D/3D RT models are generally applied to an axis-symmetric heliosphere without the distortion effect of the helio-latitudinal  variation of the solar wind parameters or of an oblique local interstellar magnetic field (LIMF). The presence of such a magnetic field is actually well-established with clear signatures predicted by models and confirmed by observations and {\it in situ} measurements of plasma and field distributions in the outer heliosphere \citep{rat98,ben00,lal05,sto05,sto08,sch09,hee11,mcc12,ben13,zir16}. However, with the smoothing effect of the charge exchange between neutrals and ions \citep{izm05,woo14}, the impact of the LIMF on the neutrals distribution and consequently on the sky background \lya\, radiation field is not too trivial to evaluate.

\subsection{Local radiation transfer approach: A first application}
In contrast to the global RT technique, another distinct approach is to consider the differential aspect of the radiation transfer equation, a technique that allows direct inversion of data obtained at locations that are close in space. This is indeed the case of sky background \lya\, maps obtained by Voyager UVS along their trajectory out to $\sim 35$\,AU from the Sun \citep{puy97,puy98}. Hereafter, the differential concept is called the local RT model. The implementation of this technique is fully based on the invariance principle to relate different maps of the sky background radiation field to recover the physical properties of the gas layer embedded between two UVS measurements \citep{amb58,puy97}. The only limitation of this technique is the requirement of accessing the radiation field that must be observed on both sides of a layer, information that is only available through 1996 (the date at which the scan platform of UVS was stopped). This limitation explains why \citet{puy98} inverted the Voyager UVS sky maps to derive the H\,I neutral density distribution up to $\sim 35$\,AU from the Sun (position of V1 in 1996). { Interestingly, \citet{puy98}  extrapolated the Voyager-based HI density measurements to infinity using the interface model of \citet{bar93} to find a local interstellar cloud (LIC) density of $\sim 0.24\pm0.05$\,${\rm cm^{-3}}$ that is consistent with most recent findings about the outer heliosphere and the LIC H\,I number density \citep{kat16}, including the IPH distribution used in the ISSI13 calibration, yet without requiring any revision of the Voyager UVS calibration}. 

\subsection{Local radiation transfer approach and the Fermi glow: Another application}
{In the following, we discuss the application of the local RT technique to the inversion of the sky background \lya\ emission excess, a feature detected by UVS deep in the heliosphere \citep{que95}. For reference, two explanations are competing to account for the observed excess: the heliosphere origin source (Fermi glow as reported in \citet{ben00}) and the galactic source origin \citep{lal11}.  Because the Fermi glow model is controversial \citep{que06,que13}, we dedicate section 4.3.1 to address all theoretical and instrumental issues, and confirm the interpretation reported in \citet{ben00}. With that result in hand, we compare the heliospheric to galactic interpretation of the origin of the \lya\, excess, emphasizing the importance of the UVS calibration that should not be arbitrarily decided (section 4.3.2).

\subsubsection{Fermi glow: Reassessment}
In 1993, the Voyager UVS team began a campaign scanning the sky background over great circles that intersect the upwind and anti-solar directions} \citep{que95}. The main goal of that campaign was to discover any asymmetry in the sky background radiation field that may be caused by the hydrogen wall near the heliopause \citep{bar93}. An excess, showing a brightness maximum close to the upwind direction, was indeed derived by \citep{que95}, yet their global RT model of solar photons scattering in a sophisticated H\,I model (accounting for the hydrogen wall) failed to properly reproduce the observations \citep{que95, que13}. In contrast, \citet{ben00} proposed an alternate interpretation related to the Fermi scattering of \lya\, photons across the hydrogen wall, while \citet{lal11} proposed a galactic origin for the observed excess. In \citet{ben00}, the idea is principally based on the spatial distribution of the hydrogen atoms  produced by charge exchanges with ions slowing down when approaching the stagnation interface \citep{bar93}. The model could fit the observed \lya\, distribution, showing that the heliosphere nose deviates $\sim 12^{\circ}$ from the upwind direction, a deviation that was associated with the presence of an oblique LIMF. Interestingly, the orientation $\sim 40^{\circ}$ of the LIMF thus far derived by \citep{ben00} nicely fits the orientation $\sim 39-42^{\circ}$ obtained by many independent studies that use Voyager plasma observations of the termination shock, IBEX ribbon, and/or the interpretation of the SOHO/SWAN hydrogen deflection plane \citep{sch09,pog10,hee11,ben12,ben13,zir16}. In addition, the derived $\sim 12^{\circ}$ deflection of the heliosphere's nose  is consistent with the $\sim 9^{\circ}$ asymmetry estimated from the 2-3 kHz heliospheric radiation \citep{fus13}, and with the constraint on the heliotail deflection from HST \lya\, absorption observations toward nearby stars \citep{woo14}.

Contrary to many misinterpretations of the model \citep{que06,que13}, the information on the LIMF orientation in  \citet{ben00} was deduced from the Voyager UVS \lya\, excess and not from the HST/GHRS spectrum; the latter was only used to support the idea of that kind of scattering across a differential velocity field. In addition, \citep{ben00} admittedly rejected the classical optically thick model of the Fermi process and clearly provided the expression that better applies to moderate opacities that occur in the heliosphere \citep{bin98,ben00}. For that reason, we do not see the logic in attaching the optically thick version of the Fermi process to \citet{ben00} when the latter clearly rejected it (e.g., their page 927 and Eq. (1) ).

One possible way to explain how the Fermi process, as applied by \citet{ben00}, worked well in fitting the sky \lya\, excess and deriving the correct information about the LIMF orientation is that the model was applied as a local RT technique. Indeed, what matters is that the proposed process quantitatively describes the scattering events of photons on atoms that are not symmetrically distributed over space, which helps attach the observed asymmetry of the \lya\, excess to the density and velocity field asymmetries and, consequently, to the LIMF obliquity. Looking back to the Voyager UVS observations, it was possible to recover the LIMF orientation from the Voyager UVS excess invoking any local RT model of photon scattering on the asymmetric hydrogen distribution as caused by the LIMF orientation and the charge exchange process near the interface. For that reason, \citet{rat08} could recover the same orientation $\sim 40^{\circ}$ of the LISM magnetic field using the tilt of the peak of the plasma distribution away from the upwind direction as a direct measure of the peak of the density of neutrals that are at the origin of the \lya\, excess observed by Voyager UVS. In addition, the same study provided the first 3D direction of the LIMF that is now fully confirmed by Voyager plasma and IBEX observations \citep{ben12,ben13}. { Hereafter, we make reference to the heliospheric origin for both the Fermi glow interpretation that is related to the velocity field asymmetry \citep{ben00} or to the density asymmetry as reported in \citet{rat08}}.

Apart from the model concept, two additional arguments were commonly used to question the existence of the Fermi feature in the HST/GHRS spectrum reported in \citep{ben00}. The first claim was that the feature was spectrally coincident with the D \lya\, line of the Earth's geocorona \citep{que13}. On the caption of their Figure 2, \citet{ben00} clearly state that the GHRS Echelle mode with the large science aperture could not resolve both the Fermi and the telluric D \lya\, lines, and that the latter was properly subtracted using brightness levels estimated from independent sky observation obtained with the same instrument toward Mars {and in the same conditions of solar flux and HST observing geometry \citep{kra98}. For reference, the observations were scheduled on January 20, 1997 for Mars and both April 4, 1994 and March 25, 1995 for the IPH observations reported in \citep{ben00}. In addition, far UV observations are usually scheduled during Earth's nighttime of the HST orbit, so the observing geometry with respect to Earth remains unchanged. According to the lasp.colorado/lisird/lya/ site, for the 1997 Mars observation, the solar \lya\, irradiance was $\sim3.6\times10^{11}\, {\rm photons\, cm^{-2}\, s^{-1}}$. For the 1994 IPH, the irradiance was $\sim3.8\times10^{11}\, {\rm photons\, cm^{-2}\, s^{-1}}$, and for the 1995 IPH, it was $\sim3.9\times10^{11}\, {\rm photons\, cm^{-2}\, s^{-1}}$. This means that no noticeable difference exists between the solar \lya\, fluxes for those three dates (solar minimum activity). {To the best of our knowledge, no one questions the D \lya\, detection of Mars ($\sim20\pm6$\,R as reported in \citet{kra98}), which was obtained with the same instrument and in the same conditions of solar flux and observing geometry as for the IPH observations. For reference, the extra emission feature reported in \citet{ben00} has a brightness of $\sim 38\pm 6$\,R. After subtraction of the telluric D line ($\sim21\pm6$\,R), we obtain a Fermi line brightness of $\sim 17\pm6$\,R ($\sim 3\sigma$  detection).}

\begin{figure}[h]
\centering
\hspace*{-0.15in}
\includegraphics[width=9cm]{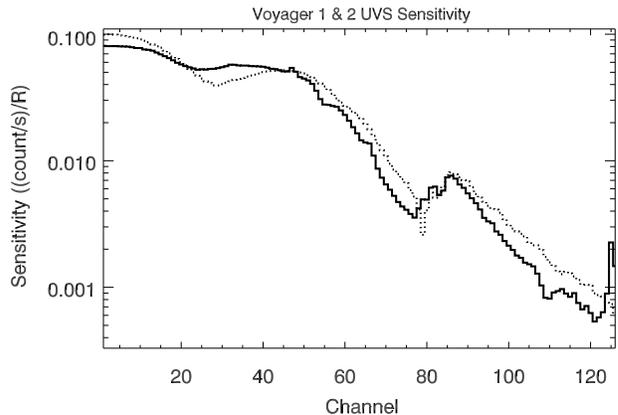}
\caption{ Reference sensitivity curves for V1 (solid)  and V2 (dash) UVS.}
\end{figure}

The second argument against the Fermi process is based on the non-detection of the Fermi feature in subsequent observations made with the Space Telescope Imaging Spectrometer (STIS) onboard the HST in the upwind direction \citep{que13}. { Here, it is necessary to remind that \citet{ben00} were cautious, clearly stressing that the interpretation of the GHSR data in terms of Fermi line cannot be conclusive until high-resolution detection of the extra emission is obtained}. Generally, a non-detection means that either the spectral feature does not exist or the instrument does not have the capability. We argue that the STIS Echelle 140H grating has no capability to detect faint emissions. In the following, we propose two main reasons to explain the non-detection of faint emissions by HST/STIS. First,  the STIS 140H Echelle grating does not have the sensitivity to detect faint emissions. Indeed, using the HST/STIS exposure time calculator (ETC) for that mode, it is not difficult to find that an extended source of $\sim20$\,R ($\sim7.0\times 10^{-16}\,{\rm ergs\,cm^{-2}\,s^{-1}\,arcsec^{-2}}$) produces a count rate that represents less than one percent of the background count rate (sky and dark current). This means that, even using an extended slit to gather more light, the signal will be dominated by noise, and the detection of such a faint source will be impossible with the STIS/E140H grating. In contrast, the GHRS has proven in the past its unique capability in counting events rapidly at a rate of 20,000 counts per second per diode,  compared to a local count rate of $\sim 50$ per second per pixel for STIS (HST/STIS handbook, 2006). This underscores the fact that GHRS helped detect faint emissions, particularly the telluric deuterium \,lya\, emision line \citep{kra98}. In contrast, the same emission line has never been detected with STIS. The first author participated in many campaigns to the study the sky background  \lya\, line profile in the upwind direction obtained with HST/STIS since 2001, and the D \lya\, line was detected in none of them \citep{vin11,vin14}. All those facts, taken together,  explain why the HST/STIS E140H mode is not capable of detecting the Fermi or any faint emission line.

\subsubsection{Heliospheric or galactic origin? The UVS calibration answer}

There are two competing explanations for the sky background \lya\, excess observed by UVS: the heliosphere origin source (our model) and the galactic source origin \citep{lal11}. To support the galactic origin, \citet{lal11} proposed a potential correlation between the UVS \lya\, distribution and H$\alpha$\, emission from the H II region over the sky. The approach is interesting because if the observed emission has a galactic origin, then the correlation should be strong at any time and any place in space, particularly if using two distinct spacecraft (V1 \& V2 in the present case). The problem is that when looking back at Figure 4 in \citet{lal11}, {it easy to verify that the reported correlation between two \lya\, scans (out of four available) and the H$\alpha$\, distribution over the sky is poor. In addition to the spatial distribution problem, the issue of the brightness level of the \lya\, excess received little attention. Indeed, \citet{lal11} used a UVS calibration of  70 R/count $s^{-1}$ and 80 R/count $s^{-1}$ respectively for V1 and V2 to obtain a \lya\, excess of 3-4\, R, a brightness that corresponds to a signal level in the range 0.02-0.05 count $s^{-1}$. In contrast, \citet{que95} derived a brightness of 10-15 R based on signal level of $\sim 0.1$ count $s^{-1}$ and using a UVS calibration that is distinct from \citet{lal11} and from \citet{que13}. For reference, using the ISSI13 calibration, the observed \lya\, emission excess reported in \citet{que95} is three times the values reported in \citet{lal11}, while using the VOY92 calibration the emission ($\sim 17-22$\, R)  is at least 4 times that value. { This means that the \lya\, excess brightness exceeds the estimation made by \citet{blu89} and the upper theoretical limit of $\sim 10$\, R derived by \citet{tho76} for our galaxy based the same H$\alpha$\, distribution invoked by \citet{lal11} for their proposed correlation.} 

In contrast, the Fermi glow brightness $\sim 17\pm6$\,R observed by HST/GHRS  is consistent with the UVS \lya\, excess of $\sim 17-22$\, R derived with the original calibration (e.g., section 4.3.1). In addition, a heliospheric origin of the UVS \lya\, excess \citep{ben00,rat08} leads to properties of the LISM magnetic field orientation that are now accurately confirmed by SOHO/SWAN, Voyager plasma, Voyager radio emission, IBEX ribbon, and HST \lya\, absorption measurements \citep{lal05,sch09,hee11,ben12,ben13,fus13,mcc12,woo14,zan15,zir16}. Furthermore, recent sophisticated 3D MHD-kinetic and RT modeling of the outer heliosphere observations obtained by UVS during the 1993-2003 period led to the important result that a high hydrogen density is required to fit the data \citep{kat16}, a finding that is consistent with the high density level derived by \citet{puy97,puy98} with the original UVS calibration, and confirmed here for the inner heliosphere ($< 10$\, AU). 

All arguments discussed above question the galactic origin of the UVS \lya\, excess, and support its heliospheric origin, making of the Voyager UVS the first instruments that detected the distortion of the heliosphere, and consequently the presence of an ambient oblique interstellar magnetic field in the vicinity of our solar system \citep{ben00}.}

\section{Conclusions: a first step} 
During the last two decades, the Voyager 1 and 2 ultraviolet spectrometers harvest covered EUV \& FUV  observations of the outer planets and their satellites, heliosphere sky-background in situ measurements (\lya\,, \lyb\,, He 58.4\,nm), and stellar spectrophotometry. {Recently, \citet{que13} used sophisticated modeling of the sky background emission in the outer heliosphere to revise the UVS instruments calibration by no less than a factor 243\% for V1 and 156\% for V2 compared to the post-Jupiter encounter calibration \citep{hal92}}. Because a strong modification of the UVS calibration has a much broader impact on many other topics than just the heliosphere problem, a reassessment was required.

Here, we use simultaneous Voyager UVS (encounters) and IUE (remote) observations of the dayglow of Saturn to reevaluate the V1 and V2 UVS calibration. For that purpose, we first use HST/GHRS archive high-resolution observation of the Saturn \lya\, line profile to derive that the planetary emission line width $\sim 0.01$\,nm is large enough to show an absorption signature due to the moving IPH atoms between Saturn and Earth for all possible ranges of the Doppler-shifted absorption. With the Saturn \lya\, line profile in hand, we have compared V1 and V2 brightness of Saturn to the brightness observed simultaneously by IUE at the Earth orbit using both the old and new calibration of the Voyager instruments. As shown in Tables 2 \& 3, with the recent ISSI13 calibration of V1 UVS, the Saturn emissions brightness on 1980 is below levels measured by IUE at the Earth orbit, which is fully in contradiction with the IPH model used by \citet{que13} to derive their calibration. Even were it possible for the interplanetary medium to be empty, the lower intensities derived for V1 are not consistent with emissions brightness observed at Earth unless the IUE calibration itself is revised. Because the IUE sensitivity is well-established \citep{boh90}, matching UVS ISSI13 with IUE observations would require a strong changes in the IUE sensitivity over one year between 1980 (V1) and 1981 (V2), which have never been reported in any other IUE observations. Such strong variations on short time scales are thus unrealistic, thereby casting doubt on the ISSI13 revision of the V1 and V2 UVS calibration. In addition, for the ISSI13 calibration of V2 UVS, the difference between the V2 and IUE measurement is less than the uncertainty related to the UVS measurements, which led to  a marginal solution at the limit of the noise level with a low hydrogen density at 10 AU,  almost a factor of 2 below the hydrogen density assumed in the model used to derive the ISSI13 calibration (e.g., Table 3).  {Finally, the comparison of our radiation transfer modeling of the  HeI 58.4\,nm airglow of Jupiter measured by V1 \& V2  to the  Galileo probe measurement of the helium abundance {\it in situ}  the Jovian atmosphere leads to the same diagnostic  on the validity of the UVS original calibration but at distinct channels covering the diffuse HeI 58.4\,nm emission}.

With the strong evidence reported here for the stability of their calibration, both Voyager UVS almost certainly become one of the most stable EUV/FUV spectrographs of the history of space exploration. The final uncertainty of 30\% derived here on the UVS sensitivity includes the nominal inflight uncertainty of 20\% thus far reported by the UVS team \citep{hol82}, in addition to the potential variability of the planetary airglow used in the calibration process, particularly during the comparison with the IUE observations  \citep{puy97}. For reference, we show the final V1 \& V2 UVS sensitivity curves for extended sources in Figure 7.

One possible way to explain the inadequacy of the ISSI13 calibration is related to the technique used to solve the complex RT problem in the heliosphere. While a differential local approach of the sky \lya\, background modeling appears as a self-consistent inversion technique to derive the interplanetary medium composition without requiring any revision of instruments calibration, global models (like the one used in ISSI13) that include the whole heliosphere and its complex interaction with the LIC seem to fail to capture the key feature of the physics of the problem, which causes unrealistic modification of the sensitivity of Voyager UVS instruments. {With this in mind, we reconsider the local RT inversion of UVS sky-background \lya\, maps to confirm the high H\,I density level reported two decades ago by \citet{puy97,puy98} for the inner heliosphere, a finding that is now supported by sophisticated 3D RT interpretation of UVS measurements obtained in the outer heliosphere \citep{kat16}. In addition, we provide evidence that the interpretation of the \lya\, emission excess, a feature detected since 1995 by Voyager UVS deep in the heliosphere, is not consistent with a galactic origin as reported in \citet{lal11}. In contrast, our theoretical and experimental reassessment of the Fermi glow { and, more generally, of the heliospheric} interpretation of the same \lya\, excess, confirms our finding on the heliosphere distortion and the corresponding local interstellar magnetic field's obliquity ($\sim 40^{\circ}$ from upwind) \citep{ben00}, without requiring any revision of the Voyager UVS calibration. The above arguments make both Voyager UVS the first instruments that detected two decades ago the distortion of the heliosphere and, consequently, the presence of the oblique interstellar magnetic field in the neighborhood of our solar system \citep{ben00,ben12}}.

To build on the rich heritage from both Voyager UVS and past space missions, our study points to the need in the future for fine Doppler-shift measurements and faint emissions detection (probably including polarimetry) in different locations of the heliosphere so that direct inversion techniques can be applied in order to directly access the microphysical processes that drive the instant shape and composition of the heliosphere that is forced by the magnetized plasmas from solar wind and the local interstellar medium. Future deep space missions should thus include robust UV capabilities that make use of sensitive, high-resolution technology that will make it possible to attain the highest throughput for extended light sources \citep{ben05,har05}.

\acknowledgements
L.B.-J acknowledges support from CNES, Universit\'e Pierre et Marie Curie (UPMC) and the Centre National de la Recherche Scientifique (CNRS) in France. He also acknowledges his current appointment as a Visiting Research Scientist at the Lunar and Planetary Laboratory of the University of Arizona. Authors are grateful to Dr. G. Ballester (LPL) for helping derive the exact pointing of the GHRS observations. They also thank anonymous referees for very helpful suggestions that improved the manuscript.

\end{document}